\documentclass[a4paper]{jpconf}
\usepackage{graphicx}

\newcommand{\be}{\begin{equation}}
\newcommand{\ee}{\end{equation}}
\newcommand{\bea}{\begin{eqnarray}}
\newcommand{\eea}{\end{eqnarray}}
\newcommand{\nn}{\nonumber}

\newcommand{\ep}{\epsilon}
\newcommand{\de}{\delta}
\newcommand{\De}{\Delta}
\newcommand{\om}{\omega}

\newcommand{\ov}{\overline}

\newcommand{\op}{\overline\Pi}

\newcommand{\vk}{\vec k}

\newcommand{\vq}{\vec q}

\newcommand{\mn}{{\mu\nu}}

\begin{document}

\setcounter{page}{1}

\title{In-medium vector mesons and low mass lepton pairs from heavy ion
collisions}
\author{Sourav Sarkar and Sabyasachi Ghosh}

\address{Theoretical Physics Division, Variable Energy Cyclotron Centre,\\
1/AF, Bidhannagar, 
Kolkata 700064, India}

\ead{sourav@vecc.gov.in}

\begin{abstract}
 
The $\rho$ and $\om$ meson self-energy at finite temperature and baryon density 
have been analysed for
an exhaustive set of mesonic and baryonic loops
in the real time formulation of thermal field theory. The large
enhancement of spectral strength below the nominal $\rho$ mass is seen to cause a
substantial enhancement in dilepton pair yield in this mass region. The integrated yield after 
space-time evolution using relativistic hydrodynamics with quark gluon plasma 
in the initial state leads to a very good agreement with the experimental data from In-In
collisions obtained by the NA60 collaboration.

\end{abstract}

\section{Introduction}

Colliding heavy ions at ultra-relativistic energies is the only way to produce and study
bulk properties of strongly interacting matter. Systematic efforts, both theoretical and
experimental over the last few decades~\cite{jaipurbook,cbmbook} have addressed various facets of the
thermodynamics of the underlying theory $-$ QCD.  
The spectra of
hadrons emitted after freeze-out of the fireball produced in heavy ion collisions have
provided us with a wealth of information. This includes the recent discovery made by
studying the elliptic flow of hadrons that the quark gluon plasma (QGP) produced in Au+Au
collisions at RHIC actually behaves as a strongly interacting fluid~\cite{csernai} as opposed to a
asymptotically free gas of quarks and gluons.
However, by virtue of the fact that electromagnetic probes (real photons and dileptons)
are emitted all through the lifetime of
the fireball coupled with their low rescattering probability make them penetrating probes
capable of mapping the space-time history of the collision~\cite{annals}.
Lepton pairs with both invariant mass and transverse momentum information  are 
in fact preferable to real photons. Large mass pairs produced by Drell-Yan process and from the
decays of heavy quarkonia are emitted early whereas pairs with low invariant 
mass radiated from thermal hadronic matter and
Dalitz decays of hadrons are produced late in the collision. The invariant mass spectra of dileptons 
thus carry time information as displayed explicitly in~\cite{payal} using the invariant mass
dependence of the elliptic flow of lepton pairs.

The rate of production of thermal dileptons is proportional to the two-point correlator of
vector currents~\cite{toimela}. In the low invariant mass region which is dominated by
lepton pairs produced during the later stages of the collision, dilepton emission 
takes place due to the decay of vector mesons. 
Consequently, the spectral properties of vector mesons, the $\rho$ meson in particular
has been a subject of intense discussion~\cite{rappadv,annals,cbmbook,leopoldrev}.
We find that it is only for the $\pi-\pi$ loop that one calculates the thermal 
self-energy loop for the vector mesons directly. In the case of other loops
typically involving one heavy and one light particle or both heavy particles 
one uses in general either the virial formula or the Lindhard function. 

The sources modifying the free propagation
of a particle find a unified description in terms of contributions from the 
branch cuts of the self energy function. In addition to the unitary cut present 
already in vacuum, the thermal amplitude generates a new
cut, the so called the Landau cut which provides the effect of collisions 
with the surrounding particles in the medium. This formalism was applied 
to obtain the $\rho$ self-energy in hot mesonic~\cite{ghosh1} and
baryonic~\cite{ghoshnpa} matter considering an exhaustive set of 
one-loop diagrams.  
The framework of real time 
thermal field theory that we use, enables us to evaluate 
the imaginary part 
of the self-energy from the branch cuts for real and positive values of
energy and momentum without having to resort to analytic continuation
as in the imaginary time approach.
To evaluate the baryonic loops we work with the full relativistic baryon propagator
in which baryons and anti-baryons manifestly appear on an equal footing. Thus
the contributions from all the 
singularities in the self-energy function including the distant ones coming from 
the unitary cut
of the loops involving heavy baryons are also included. 
These are not considered in the Lindhard function approach but can contribute appreciably to the real part
of the $\rho$ meson self-energy as shown~\cite{ghosh3} in the case of a $N\Delta$
loop.

The broadening of the vector meson spectral functions leads to an enhancement of lepton pair
production in the invariant mass region below the $\rho$ peak. The effect of the evolving
matter is handled by relativistic hydrodynamics. The integrated yield is seen to agree
very well with the NA60 data~\cite{sanja} from In-In collisions at 17.3 AGeV.

The article is organised as follows. We will begin with a short derivation
of the the dilepton emission 
rate in terms of the current correlation function in section 2. The relation to the spectral
function of vector mesons is specified in section 3. This will be followed by a discussion on $\rho$ and 
$\om$ self-energies in hot and
dense matter in section 4. In section 5 a brief account of the space-time 
evolution and initial conditions will be
provided followed by the dilepton invariant mass spectra. We will end 
with a summary in section 6.

\section{Dilepton emission rate and the current correlation function}

Let us consider an initial state $|I\rangle$ 
which goes to a final state $|F\rangle$ 
producing a lepton pair $l^+l^-$ with momenta $p_1$ and $p_2$ respectively.
The dilepton multiplicity thermally averaged over initial states
is given by~\cite{weldon}
\be 
N=\sum_{I}\sum_{F}|\langle F,l^+l^-|e^{i\int {\cal L}_{int}d^4x}| I\rangle|^2 \frac{e^{-\beta E_I}}{Z}\frac{d^3p_1}{(2\pi)^32E_1}\frac{d^3p_2}{(2\pi)^32E_2}
\ee
where $Z=Tr[e^{-\beta H}]$ and ${\cal L}_{int}=
e\overline{\psi}_l(x)\gamma_{\mu}\psi_l(x)A^{\mu}(x)+eJ_\mu^{h}(x)A^{\mu}(x)$ in
which $\psi_l(x)$ is the lepton field operator 
and $J_\mu^{h}(x)$ is the electromagnetic current of hadrons.
Following~\cite{toimela,weldon,annals} this expression can be put in the form 
\be
\frac{dN}{d^4xd^4q}=\frac{e^4L(q^2)}{3(2\pi)^5q^4}e^{-\beta q_0} W^{>}_{\mu\nu}(q)(q^{\mu}
q^{\nu}-q^2g^{\mu\nu})
\ee
where, $W^{>}_{\mu\nu}=\int d^4x\ e^{iq\cdot x}\langle J^h_{\mu}(x)J^h_{\nu}(0)
\rangle_{\beta}$ is the Fourier transform of the thermal
expectation value of the two-point correlator of the hadronic currents
and $L(q^2)=(1+\frac{2m^2_l}{q^2})\sqrt{1-\frac{4m^2_l}{q^2}}$. 
We now define $W^{<}_{\mu\nu}$ by interchanging the order of the currents
getting $W^{<}_{\mu\nu}=e^{-\beta q_0}W^{>}_{\mu\nu}$. Using these to define the
commutator 
$W_{\mu\nu}=W^{>}_{\mu\nu}-W^{<}_{\mu\nu}=\int d^4x\ e^{iq\cdot x}\langle 
[J^h_{\mu}(x),J^h_{\nu}(0)]\rangle_{\beta}$ we finally have
\be
\frac{dN}{d^4xd^4q}=-\frac{\alpha^2}{6\pi^3}\frac{g^{\mu\nu}}{q^2}L(q^2)
f_{BE}(q_0) W_{\mu\nu}(q_0,\vq)
\label{rate}
\ee
for conserved hadronic currents.
This expression appears in different forms in the literature.
Replacing $f_{BE}(q_0)W_{\mu\nu}$ in (\ref{rate}) by 
(i)$W^{M}_{\mu\nu}=\int d^4x\ e^{iq\cdot x}\langle J^h_{\mu}(x)J^h_{\nu}(0)\rangle_{\beta}$
yields the expression in~\cite{toimela}, by 
(ii) $2{\rm {Im}} W^{R}_{\mu\nu}f_{BE}(q_0)$
where $W^{R}_{\mu\nu}=i\int d^4x\ e^{iq\cdot x}\theta(t-t')
\langle [J^h_{\mu}(x),J^h_{\nu}(0)]\rangle_{\beta}$ yields 
the rate in~\cite{rapp_chiral}  and by 
(iii) $2{\rm {Im}} W^{T}_{\mu\nu}/(1+e^{\beta q_0})$ where
$W^{T}_{\mu\nu}=i\int d^4x\ e^{iq\cdot x}\langle TJ^h_{\mu}(x)J^h_{\nu}(0)
\rangle_{\beta}$ gives the rate in~\cite{zahed}.

The rate given by eq.~(\ref{rate}) is to leading order in electromagnetic 
interactions but exact to all orders
in the strong coupling encoded in the current correlator $W_{\mu\nu}$. The $q^2$ in 
the denominator indicates the exchange of a single virtual photon and 
the Bose distribution implies the thermal weight of the source.

\section{Current correlator and the spectral function of vector mesons} 

Thermal field theory is the appropriate framework to carry out 
perturbative calculations in the medium.
In the real time version of this formalism two point functions assume
a $2\times 2$ matrix structure on account
of the shape of the contour in the complex time plane~\cite{realtime,bellac}. 
It is convenient to begin with the quantity,
\be
T^{ab}_{\mu\nu}=i\int d^4x e^{iq\cdot x}\langle T_cJ^h_{\mu}(x)J^h_{\nu}(0)\rangle_{\beta}
\label{Tab}
\ee
where $T_c$ denotes ordering along the contour and $a, b$ are the thermal indices
which take values 1 and 2.
This quantity can be diagonalised by means of a matrix $U$ so that 
\be
T^{ab}_{\mn}=U\left(\begin{array}{cc}\ov T_{\mn} & 0\\0 & -\ov T_{\mn}^*\end{array}
\right)U~;~~~~U=\left(\begin{array}{cc}\sqrt{1+n} & \sqrt{n}\\
\sqrt{n} & \sqrt{1+n}\end{array}\right)~,~~~n=\frac{1}{e^{\beta|q_0|}-1}
\label{diag}
\ee
where $\ov T_\mn$ is an analytic function.
Equating both sides it follows that this function is obtainable from any one
(thermal) component of  $T^{ab}_{\mn}$. It is related e.g.
to the 11-component as,
\be
{\rm Re} \ov T_{\mn}(q_0,\vq)={\rm Re} T^{11}_{\mn}(q_0,\vq)~;~~
{\rm Im} \ov T_{\mn}(q_0,\vq)=\coth(\frac{\beta |q_0|}{2})
{\rm Im} T^{11}_{\mn}(q_0,\vq)~.
\label{t11}
\ee
Furthermore, $\ov T_{\mn}$ has the spectral representation~\cite{souravijp,realtime}
\be
\ov T_{\mn}(q_0,\vq)=\int\frac{dq'_0}{2\pi}\frac{W_{\mu\nu}(q_0',\vq)}{q'_0-q_0-i\eta\epsilon(q_0)}
\ee
which immediately leads to
\be
W_{\mu\nu}(q_0,\vq)=2\epsilon(q_0) {\rm Im} \ov T_{\mn}(q_0,\vq)~.
\label{W_T}
\ee
In the QGP where quarks and gluons are the relevant degrees of freedom,
the time ordered correlation function $T^{11}_{\mn}$ can be directly evaluated 
by writing the hadron current in terms of quarks of flavour $f$ i.e.
$J_\mu^{h}=\sum_{f}e_f \overline{\psi}_f\gamma_{\mu}\psi_f$. To leading order
we obtain using relations (\ref{t11}) and (\ref{W_T}),
\be
g^{\mn}W_{\mu\nu}=-\frac{3q^2}{2\pi}\sum_{f}e^2_f(1-\frac{4m^2_q}{q^2})~.
\ee
The rate in this case corresponds to dilepton production due to process 
$q\overline{q}\rightarrow\gamma^*\rightarrow l^+l^-$.
To obtain the rate of dilepton production from hadronic interactions 
it is convenient to break up
the quark current $J_\mu^{h}$ into parts with definite isospin
\bea
J^{h}_\mu&=&\frac{1}{2}(\bar u\gamma_\mu u -\bar d\gamma_\mu d)+ 
\frac{1}{6}(\bar u\gamma_\mu u +\bar d\gamma_\mu d) + \cdots
\nn\\
&=&J^{V}_\mu+J^{S}_\mu + \cdots
\nn\\
&=&J^{\rho}_\mu+J^{\om}_\mu/3 + \cdots
\label{v_current}
\eea
where $V$ and $S$ denote iso-vector and iso-scalar currents and
the dots denote currents comprising of quarks with strangeness 
and heavier flavours.
These currents couple to individual hadrons 
as well as multiparticle states with the same quantum numbers
and are usually labelled by the lightest meson in the corresponding
channel~\cite{shuryak}. We thus identify the isovector and isoscalar currents
with the $\rho$ and $\omega$ mesons respectively.
Using eq.~(\ref{v_current}) in eq.~(\ref{Tab}) and
neglecting possible mixing between the isospin states, we write
\be
T^{ab}_{\mn}=T^{(\rho)ab}_{\mn} + T^{(\om)ab}_{\mn}/9 + \cdots
\ee
where
\be
T^{(\rho)ab}_{\mn}=i\int d^4x e^{iq\cdot x}\langle T_cJ^\rho_{\mu}(x)J^\rho_{\nu}(0)\rangle_{\beta}
\ee
and similarly for the scalar current. 
The current commutator in the isospin basis follows as
\be
W_{\mn}=W^\rho_{\mn} + W^\om_{\mn}/9 + ...
\label{w}
\ee
The correlator of vector-isovector currents $W^\rho_{\mn}$ have in fact been 
measured~\cite{aleph} in vacuum along with the axial-vector correlator 
by studying $\tau$ decays
into even and odd number of pions. The former
is found to be dominated at lower energies 
by the prominent peak of the $\rho$ meson followed by a continuum at high
energies.
The axial correlator, on the other hand, is characterised by the broad hump of the $a_1$.
The distinctly different shape in the two spectral densities
is an experimental signature of the fact that
chiral symmetry of QCD is dynamically broken by the ground state~\cite{sarkarhad}. It is expected
that this symmetry may be restored at high temperature and/or density and will
be signalled by a complete overlap of the vector and axial-vector
correlators~\cite{kapusta}. 

\begin{figure}
\begin{center}\includegraphics[scale=0.8]{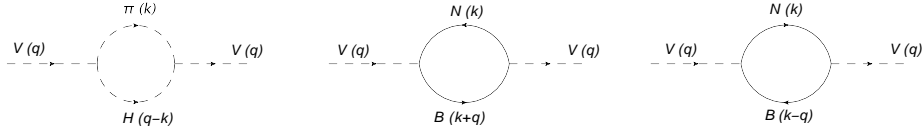}\end{center}
\caption{One-loop Feynman diagrams for $\rho$ or $\om$ self-energy involving 
mesons (first figure) and
 baryons (second and third figures). $V$ stands for the  
$\rho$ or $\om$ in the external line.
In the internal lines, $h$ stands for $\pi$, $\om$, $a_1$ and  $h_1$ mesons. 
For the baryonic loops, $N$ and $B$ indicate respectively nucleon and 
baryonic internal lines.}
\label{loop_BB}
\end{figure}

In the medium, both the pole and the continuum structure of the correlation
function gets modified~\cite{annals,vector}. We will first evaluate the 
modification of the pole
part due to the self-energy of vector mesons in the following.
Using Vector Meson Dominance the isovector and scalar currents are written 
in terms of dynamical field operators for the mesons 
allowing us to express the correlation function in terms of the exact(full) 
propagators of the vector mesons in the medium. Writing 
$J_\mu^{\rho}(x)=F_\rho m_\rho \rho_\mu(x)$ and 
$J_\mu^{\om}(x)=3F_\om m_\om \om_\mu(x)$ 
in eq.~(\ref{w}) and using eq.~(\ref{W_T}) the current commutator becomes
\be
W_{\mu\nu}=2\epsilon(q_0)F^2_\rho m^2_\rho  {\rm Im} \ov D^{\rho}_{\mn} 
+2\epsilon(q_0)F^2_\om m^2_\om {\rm Im} \ov D^{\om}_{\mn}
+ \cdots
\label{W_D}
\ee
where $\ov D_{\mn}$ is the diagonal element of the thermal propagator matrix  
which is a two point function
of the fields of vector mesons and is diagonalisable as in (\ref{diag}). 
The exact propagator is obtained in terms of 
the in-medium self-energies using the Dyson equation.
Following~\cite{ghoshdil} this is given by
\be
\ov D_{\mn}(q)=-\frac{P_{\mn}}{q^2-m_\rho^2-\op_t(q)}
-\frac{Q_{\mn}/q^2}{q^2-m_\rho^2-q^2\op_l(q)}-\frac{q_\mu q_\nu}{q^2m_\rho^2}
\ee
where $P_{\mn}$ and $Q_{\mn}$ are the transverse and longitudinal projections.
The imaginary part is then put in eqs.~(\ref{W_D}) and then in eq.~(\ref{rate})
to arrive at the dilepton emission rate~\cite{ghoshdil}
\be
\frac{dN}{d^4qd^4x}=\frac{\alpha^2}{\pi^3q^2}L(q^2)f_{BE}(q_0) \left[
{F^2_\rho m^2_\rho} A_\rho(q_0,\vq)
+{F^2_\om m^2_\om} A_\omega(q_0,\vq)
+\cdots\right]
\label{dilrate3}
\ee
where e.g. $A_\rho(=-g^{\mn}{\rm Im}\ov D^{\rho}_{\mn}/3)$ is given by
\be
A_\rho=-\frac{1}{3}\left[\frac{2\sum{\rm
Im}\Pi^R_t}{(q^2-m_\rho^2-\sum\mathrm{Re}\Pi^R_t)^2
+(\sum{\rm Im}\Pi^{R}_t)^2}+\frac{q^2\sum{\rm Im}\Pi^R_l}
{(q^2-m_\rho^2-q^2\sum\mathrm{Re}\Pi^R_l)^2
+q^4(\sum{\rm Im}\Pi^{R}_l)^2}\right]
\label{spdef}
\ee
the sum running over all mesonic and baryonic loops.
Thus, the dilepton emission rate in the present scenario actually boils down 
to the evaluation of the self energy graphs (shown in Fig.~\ref{loop_BB}). 
The self-energy is also a $2\times 2$ matrix and is diagonalisable by the matrix
$U^{-1}$. The real and imaginary parts of the self energy function
can then be obtained from the 11-component 
as~\cite{bellac,ghoshnpa}
\bea
&&{\rm Re}\Pi^R_{\mn}(q_0,\vq)={\rm Re}\ov\Pi_{\mn}(q_0,\vq)={\rm Re}\Pi^{11}_{\mn}(q_0,\vq)\nonumber\\
&&{\rm Im}\Pi^R_{\mn}(q_0,\vq)=\ep(q_0){\rm Im}\ov\Pi_\mn(q_0,\vq)=\tanh(\beta q_0/2){\rm Im}
\Pi^{11}_{\mn}(q_0,\vq)
\label{ret}
\eea
where $\Pi^R$ denotes the retarded self-energy. 

As indicated earlier, coupling of the hadronic current to multiparticle states
gives rise to a continuum structure in the current correlation function $W^\mn$. Following Shuryak~\cite{shuryak}
we take a parametrised form for this contribution and augment 
the dilepton emission rate with
\be
\frac{dN}{d^4qd^4x}=\frac{\alpha^2}{\pi^3}L(q^2)f_{BE}(q_0) \sum_{V=\rho,\om}A^{\rm cont}_V
\label{cotineum}.
\ee
where 
\be
A_\rho^{\rm cont}=\frac{1}{8\pi}\left(1+\frac{\alpha_s}{\pi}\right)
\frac{1}{1+\exp(\omega_0-q_0)/\delta}
\ee
with $\om_0 = 1.3, 1.1$ GeV for $\rho, \om$ and $\delta = 0.2$ for both $\rho$
and $\om$. The continuum contribution for the $\om$ contains an
additional factor of $\frac{1}{9}$.

\section{Spectral function and self-energy of vector mesons}

Obtaining the in-medium spectral functions for the $\rho$ and $\om$ mesons 
essentially involves the evaluation of the self-energies. For the $\rho$ meson
these have been evaluated recently for mesonic~\cite{ghosh1} and
baryonic~\cite{ghoshnpa} loops. In the following we summarise these results
for the $\rho$ followed by those for the $\om$ meson.

\subsection{$\rho$ meson} 
 
The 11-component of the self-energy of vector mesons for loops containing mesons or baryons
can be generically expressed as
\be
\Pi^{11}_{\mn}(q)=i\int\frac{d^4k}{(2\pi)^4}L_{\mn}(k,q)
\Delta^{11}(k,m_k)\Delta^{11}(p,m_p) ~.
\label{eq_b}
\ee
where the term $L_{\mn}(k,q)$ contains factors from the numerators of 
the two propagators in the loop as well as from the two vertices and
\be
\Delta^{11}(k,m_k)=\frac{-1}{k^2-m_k^2+i\ep}+a2\pi i\delta(k^2-m_k^2)N_1~.
\ee
For bosons, $a=1$ and $N_1=n$ with $n(\om_k)=\frac{1}{e^{\beta\om_k}-1}$
whereas for fermions, $a=-1$ and $N_1=n_+\theta(k_0)+n_-\theta(-k_0)$ 
with $n_{\pm}(\om_k)=\frac{1}{e^{\beta(\om_k\mp\mu)}+1}$. The loop momentum $p$
in (\ref{eq_b}) for the various cases are as shown in Fig.~(\ref{loop_BB}).

We begin with mesonic loops.
The $\rho$ self-energy for four possible $\pi$-$h$ loops, 
where $h=\pi,\om,h_1,a_1$ have been evaluated.
The imaginary part of the retarded self-energy (\ref{ret}) is given by,
\bea
&&{\rm Im}\Pi^{\mn}_R (q_0,\vq)=-\pi\int\frac{d^3\vec k}{(2\pi)^3 4\om_\pi\om_h}\times\nonumber\\
&&[L^{\mn}_1\{(1+n_\pi+n_h)\de(q_0-\om_\pi-\om_h)
-(n_\pi-n_h)\de(q_0-\om_\pi+\om_h)\}\nonumber\\
&& +L^{\mn}_2 \{(n_\pi-n_h)\de(q_0+\om_\pi-\om_h)
-(1+n_\pi+n_h)\de(q_0+\om_\pi+\om_h)\}]~.
\label{im_mes}
\eea
where the  Bose distribution functions $n_\pi\equiv n(\omega_\pi)$ with
$\omega_\pi=\sqrt{\vk^2+m_\pi^2}$ and
$n_h\equiv n(\omega_h)$ with $\omega_h=\sqrt{(\vq-\vk)^2+m_h^2}$.
 $L^{\mn}_i(i=1,2)$ are the values of $L^{\mn}(k_0)$ for
$k_0=\om_\pi,-\om_\pi$ respectively 
and the vertices used in $L^{\mn}(k_0)$ have been obtained from the 
chiral Lagrangians which are 
specified in~\cite{ghosh1}. 

\begin{figure}
\centerline{\includegraphics[scale=0.6]{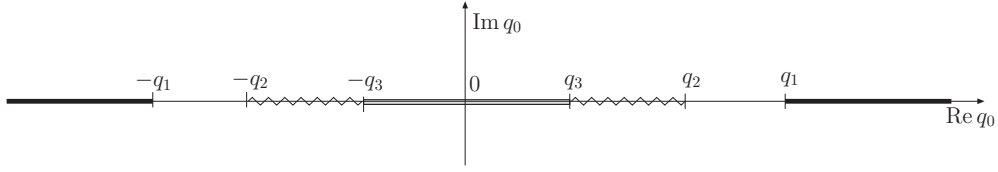}}
\caption{Branch cuts of self-energy function in $q_0$ plane for fixed $\vq$
given by $\pi h$ loop. The quantities $q_{1,2,3}$ denote the end points of cuts 
discussed in the text : $q_1=\sqrt{(m_h+m_\pi)^2+|\vq|^2}$, $q_2=\sqrt{(m_h-m_\pi)^2
+|\vq|^2}$ and $q_3=|\vq|$.}
\label{cut_diag}
\end{figure}
 
The regions, in which the four terms of eq.~(\ref{im_mes}) are non-vanishing, give
rise to cuts in the self-energy function (Fig.~\ref{cut_diag}).
 These regions are controlled by the 
respective $\de$-functions. Thus, the first and the
fourth terms are non-vanishing for $q^2 \geq (m_h +m_\pi)^2$, giving the 
unitary cut, while the second and the third are non-vanishing for 
$q^2 \leq (m_h -m_\pi)^2$, giving the so-called Landau cut. The unitary cut
 arises from the states, which can communicate with the
$\rho$. These states are, of course, the same as in vacuum, but, as we see above, 
the probabilities of their occurrence in the medium are modified by the distribution 
functions. On the other hand, the 
Landau cut appears only in medium and arises from scattering of $\rho$ with 
particles present there. We note that this contribution appears as the first 
term in the virial expansion of the self-energy function.

Integrating over the angle and restricting to the kinematic region $q_0,q^2>0$
the contribution to the imaginary part coming from the unitary cut is given by~\cite{ghosh1} 
\be
{\rm Im}\Pi^{\mn}_{R}=-\frac{1}{16\pi |\vq|}\int_{\om^-_{\pi}}^{\om^+_{\pi}}
d\om_\pi L^{\mn}_1
\{1+n(\om_{\pi})+n(q_0-\om_{\pi})\}
\label{im_U}
\ee
and that from the Landau cut is given by
\be
{\rm Im}\Pi^{\mn}_{R}=-\frac{1}{16\pi |\vq|}\int_{{\om^-_{\pi}}'}^{{\om^+_{\pi}}'} 
d\om_{\pi} L^{\mn}_2\{n(\om_{\pi})-n(q_0+\om_{\pi})\}
\label{im_L}
\ee
where
$\om_{\pi \pm}=\frac{S^2_{\pi}}{2q^2}(q_0\pm |\vq|W_{\pi})$, 
${\om_{\pi \pm}}'=\frac{S^2_{\pi}}{2q^2}(-q_0\mp |\vq|W_{\pi})$
with $W_{\pi}=\sqrt{1 -\frac{4q^2 m_\pi^2}{S^4_{\pi}}}$ and 
$S^2_{\pi}=q^2-m^2_{h}+m^2_{\pi}$.

The real part can be obtained from the imaginary part by a dispersion relation 
or can be evaluated directly from the graphs. 

Let us now turn to the baryonic loops in the $\rho$ meson self-energy.
Using $p=k-q$ for the third diagram of Fig.~\ref{loop_BB} 
in eq.~(\ref{eq_b}) the retarded  self-energy is evaluated for $NB$ loops
including all spin one-half and three-half $4-$star
resonances listed by the Particle Data Group so that $B$ stands for the 
$N^*(1520)$,  $N^*(1650)$, $N^*(1700)$, $N^*(1720)$ $\Delta(1230)$, 
$\Delta^*(1620),$ as well as the $N(940)$ itself.

As before, the imaginary part can be evaluated from the discontinuities
of the self-energy. However, the threshold for the unitary cut for the baryon
loops being far away
from the $\rho$ pole we only consider the Landau part.
The expression of the self-energy corresponding to the second and third diagrams
of Fig.~\ref{loop_BB}  
can be obtained from one another by inverting the
sign of $q$. Adding the contributions coming from the two diagrams,
the imaginary part of the self energy is given by
\be
{\rm Im}\Pi^{\mn}_R=-\frac{1}{16\pi|\vq|}\int_{\om^+_{N}}^{\om^-_{N}}d\om_N[L^{\mn}_1(-q)\{n_+(q_0+\omega_N)-n_+(\omega_N)\}
+L^{\mn}_2(q)\{n_-(q_0+\omega_N)-n_-(\omega_N)\}]
\label{im_B}
\ee
where $\om^{\pm}_{N}=\frac{S^2_{N}}{2q^2}(-q_0 \pm |\vq| W_{N})$
with $W_{N}=\sqrt{1-\frac{4q^2m_N^2}{S^4_N}}$,
$S^2_{N}=q^2-m_B^2+m_N^2$.
The factors $L^{\mn}_i(i=1,2)$ in eq.~(\ref{im_B}) are the values of $L^{\mn}(k_0)$ 
for $k_0=\om_N,-\om_N$ respectively and is obtained using gauge invariant interactions
details of which are provided in~\cite{ghoshnpa}.

The calculations described above treats the heavy mesons and baryon resonances
in the narrow width approximation. We have included  their
widths by folding with their vacuum spectral
functions as done in~\cite{manolo}.

\begin{figure}
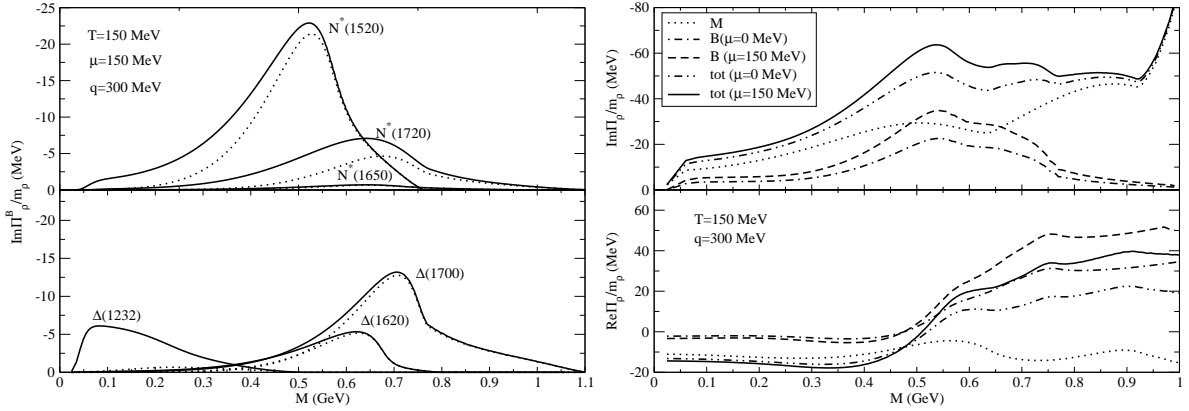

\includegraphics[scale=0.32]{im_qv_300.eps}
\includegraphics[scale=0.32]{im_re_MBt.eps}
\caption{Left panel shows the imaginary part from baryonic loops and the right 
panel shows the total contribution from meson and baryon loops.}
\label{imrepi_MBt}
\end{figure}  

\begin{figure}
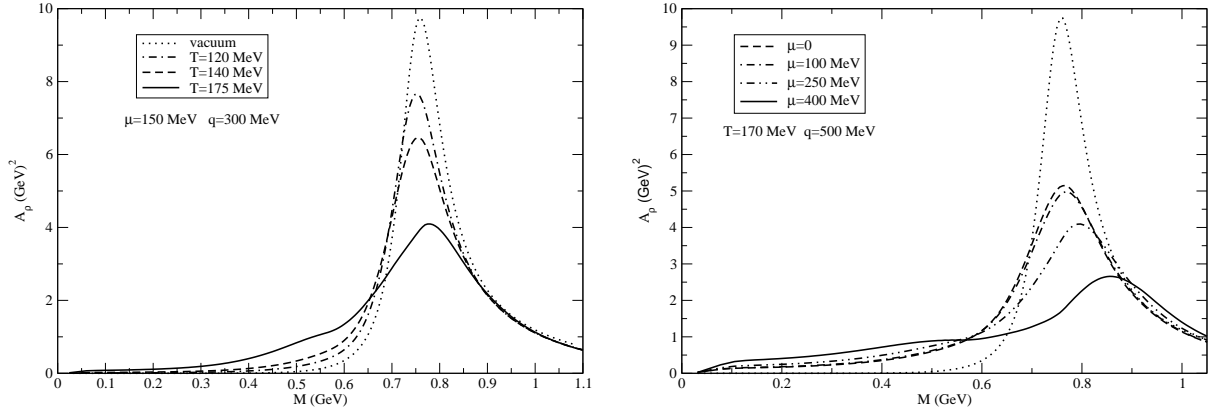

\includegraphics[scale=0.32]{spec_T.eps}~~~
\includegraphics[scale=0.32]{spec_mu.eps}
\caption{ The spectral function of the $\rho$ meson  
for (left) different values of the temperature $T$ and
(right) different values of the baryonic chemical potential ($\mu$).}
\label{spec_T_qv}
\end{figure}  

We now present the results of numerical evaluation. We plot in 
the left panel of Fig.~\ref{imrepi_MBt} the
imaginary part of the $\rho$ self-energy for baryon loops 
as a function of the invariant mass $\sqrt{q^2}\equiv M$ for $\vq=300$.
The transverse (solid line) and 
longitudinal (dashed line) components ${\rm Im}\Pi_t$ and $q^2{\rm Im}\Pi_l$ 
have been shown separately. 
The $NN^*(1520)$ loop makes the most significant contribution followed by
the $N^*(1720)$ and $\De(1700)$. 
On the right panel is plotted the spin-averaged $\rho$ self-energy 
defined by $\Pi=\frac{1}{3}(2\Pi_t+q^2\Pi_l)$
showing contributions from the baryon and meson 
loops for two values of the baryonic chemical potential. The small positive
contribution from the baryon loops to the real part is partly compensated
by the negative contributions from the meson loops. The
substantial baryon contribution at vanishing baryonic chemical potential 
reflects the importance of anti-baryons.

We now turn to the spin averaged spectral function defined in eq.~(\ref{spdef}).
First, in the left panel of Fig.\ref{spec_T_qv} we plot the spectral function at 
fixed values of the baryonic chemical potential and three-momentum for various
representative values of the temperature. 
We observe an increase of spectral strength at lower 
invariant masses resulting in broadening of the spectral function 
with increase in temperature. 
This is purely a Landau cut
contribution from the baryonic loop arising from the scattering of the 
$\rho$ from
baryons in the medium.
We then plot in the right panel of Fig.~\ref{spec_T_qv}, the spectral function 
for various values of the baryonic chemical potential for a fixed temperature.
For high values of $\mu$
we observe an almost flattened spectral density of the $\rho$.  

\subsection{$\omega$ meson}

The $\omega$ self-energy is evaluated along similar lines~\cite{ghosh_om}.
The $\omega$ meson decays mostly into three pions. Assuming this decay to proceed
via an intermediate $\rho$ meson i.e. $\omega\rightarrow \rho\pi\rightarrow 3\pi$,
the dominant contribution to the $\om$ self-energy in meson matter
can be expressed as~\cite{leupold_om}
\be
\Pi^{\mn}_{R(3\pi)}(q)=\frac{1}{N_{\rho}}\int^{(q-m_{\pi})^2}_{4m^2_{\pi}} dM^2[\Pi^{\mn}_{R(\pi\rho)}(q,M)]A_{\rho}(M)
\label{3pi_med}
\ee
where $N_{\rho}=\int^{(q-m_{\pi})^2}_{4m^2_{\pi}} dM^2 A_{\rho}(M^2)$, 
$A_{\rho}$ being the vacuum spectral function of the $\rho$.
Here $\Pi^{\mn}_{R(\pi\rho)}(q,M)$ can be obtained by
evaluating the first diagram of Fig.~(\ref{loop_BB}) with $h=\rho$. 

For $\om$ self-energy due to baryons we have evaluated $NB$
loops where $B = N^*(1440), N^*(1520), N^*(1535), N^*(1650), N^*(1720), N(940)$. The
calculation proceeds similarly as in case of the $\rho$.

\begin{figure}
\includegraphics[scale=0.32]{fig3.eps}
\includegraphics[scale=0.32]{fig8.eps}
\caption{Same as Fig.~(\ref{imrepi_MBt}) for the $\om$.} 
\label{mu0}
\end{figure} 

\begin{figure}
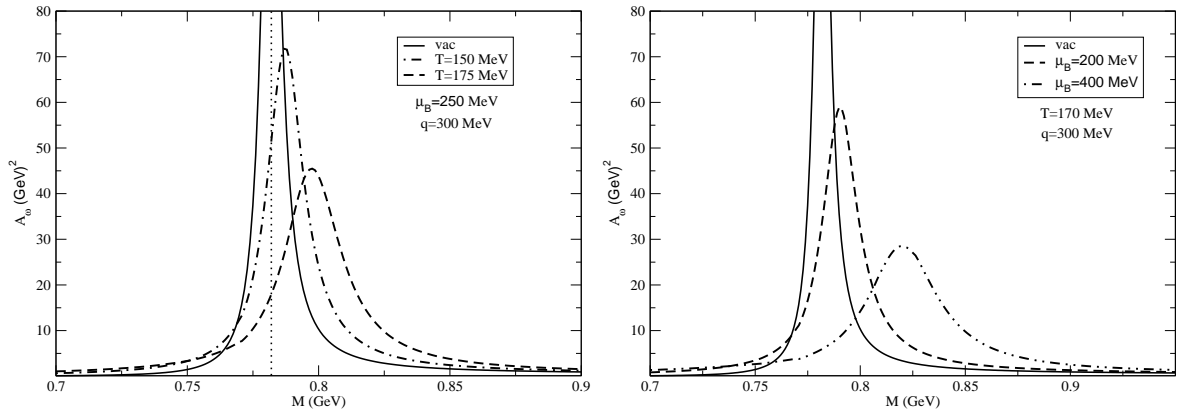

\includegraphics[scale=0.32]{fig7.eps}
\includegraphics[scale=0.32]{fig6.eps}
\caption{The spectral function of the $\om$ for different
values of $T$ (left panel) and $\mu$  (right panel).}
\label{spec}
\end{figure} 
 
Left panel of Fig.~(\ref{mu0}) shows the transverse and longitudinal components of
the imaginary part coming from the baryonic loops and the right panel shows the  
individual mesonic and baryonic 
loop contributions to the $\om$ self-energy at two different chemical
potentials. We see that the $N^*(1535)$ plays the dominating role mainly due to the
strong coupling compared to the other baryonic resonances~\cite{leupold_om}. 
The spin averaged spectral function at different
temperatures and chemical potentials are plotted in the left and right panels of 
Fig.~(\ref{spec}) respectively. We observe a slight positive shift in the peak
position.

\begin{figure}
\begin{center}\includegraphics[scale=0.32]{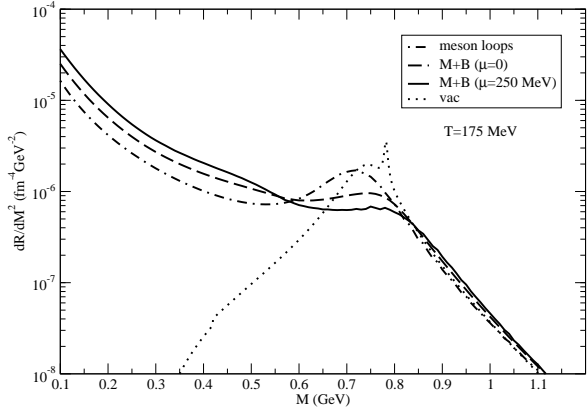}\end{center}
\caption{The lepton pair emission rate at $T=175$ MeV with and without
baryon (B) loops in addition to the meson (M) loops.}
\label{dilfig}
\end{figure}

Having obtained the in-medium spectral functions of the two most important
low-lying vector mesons, we are now in a position to evaluate the
static rate of dilepton production using eq.~(\ref{dilrate3}).
Integrating over the transverse momentum $q_T$ and rapidity $y$ of the 
electron
pairs we plot $dR/dM^2$ vs $M$ in Fig.~\ref{dilfig} for $T$=175 MeV. 
Because of the
kinematical factors multiplying the $\rho$ spectral function 
the broadening appears magnified in the dilepton
emission rate.
A significant enhancement is seen in the low mass lepton production rate
due to baryonic
loops over and above the mesonic ones shown by the dot-dashed line.
The substantial contribution from baryonic loops even for vanishing
chemical potential points to the important role played by antibaryons in thermal
equilibrium in systems created at RHIC and LHC energies. 

\section{Space time evolution and dilepton spectra in In-In collisions}

Since dileptons are produced at all stages of the collision it is necessary to
integrate the emission rates over the space-time volume from
creation to freeze-out. We assume  
that quark gluon plasma having a temperature $T_i$
is produced at an initial time $\tau_i$. Hydrodynamic expansion and cooling follows up to
a temperature $T_c$ where QGP undergoes a transition to hadronic matter. Subsequent
cooling leads to freeze-out of the fluid element into observable hadrons.
In the present work the fireball is taken to undergo an azimuthally symmetric
transverse expansion along with a boost invariant longitudinal expansion~\cite{bjorken}.
The local temperature of the fluid element and the associated flow
velocity as a function of the radial coordinate and proper time is obtained
by solving the the energy momentum conservation equation
$\partial_\mu\,T^{\mu \nu}=0$
where $T^{\mu \nu}=(\epsilon+P)u^{\mu}u^{\nu}\,+\,g^{\mu \nu}P$ is the energy
momentum tensor for ideal fluid.  This set
of equations are closed with the Equation of State (EoS); 
typically a functional relation between the pressure $P$ and 
the energy density $\epsilon$. It is a crucial input which essentially controls
the profile of expansion of the fireball. 

The initial temperature is constrained by the experimentally measured hadron multiplicity
through entropy conservation~\cite{hwa},
\be
T_i^3(b_m)\tau_i=\frac{2\pi^4}{45\zeta(3)\pi\,R_{\perp}^2 4a_k}
\langle\frac{dN}{dy}(b_m)\rangle
\label{intemp}
\ee
where $\langle dN/dy(b_m)\rangle$ is the hadron (predominantly pions) 
multiplicity
for a given centrality class with maximum impact parameter $b_m$, 
$R_{\perp}$ is the transverse dimension of the system and
$a_k$ is the degeneracy of the system created.
The initial radial velocity, $v_r(\tau_i,r)$ 
 and energy density, $\epsilon(\tau_i,r$) 
profiles are taken as~\cite{von},
$v_r(\tau_i,r)=0~~{\rm and}~~
\epsilon(\tau_i,r)={\epsilon_0}/({e^{\frac{r-R_A}{\delta}}+1})$
where the surface thickness, $\delta=0.5$ fm. 
In the present work we assume $T_c = 175$ MeV~\cite{borsanyi}. 
In a quark gluon plasma to hadronic matter transition scenario, 
we use the bag model EoS for the QGP phase and
all resonances with mass $\leq 2.5$ GeV for the hadronic gas.  
The transition region is parametrized as~\cite{hatsuda}
\be
s=f(T)s_q + (1-f(T))s_h~~{\rm with}~~
 f(T)=\frac{1}{2}(1+\mathrm {tanh}(\frac{T-T_c}{\Gamma}))~.
\ee
where $s_q$ ($s_h$) is the entropy density of the quark (hadronic) 
phase at $T_c$. 
The value of the parameter $\Gamma$ can be varied to
make the transition strong first order or continuous. 
We take $\Gamma=20$ MeV in this work. 

The ratios of various hadrons measured experimentally at 
different $\sqrt{s_{\mathrm {NN}}}$
indicate that the system formed in heavy ion collisions chemically decouple 
at a temperature ($T_{\mathrm {ch}}$) which is higher than the temperature for 
kinetic freeze-out ($T_f$)determined
by the transverse spectra of hadrons~\cite{pbm}. 
Therefore, the system remains out of chemical equilibrium
from $T_{\mathrm {ch}}$ to $T_f$.
The chemical non-equilibration affects the dilepton yields through 
(a) the emission rate through the phase space factor and 
(b) the space-time evolution of the matter via the equation of state. 
The value of the chemical potential and
its inclusion in the EoS has been taken into account 
following~\cite{hirano}.

\begin{figure}
\begin{center}
\begin{tabular}{ccc}
\includegraphics[height=5.6cm,width=7.5cm]{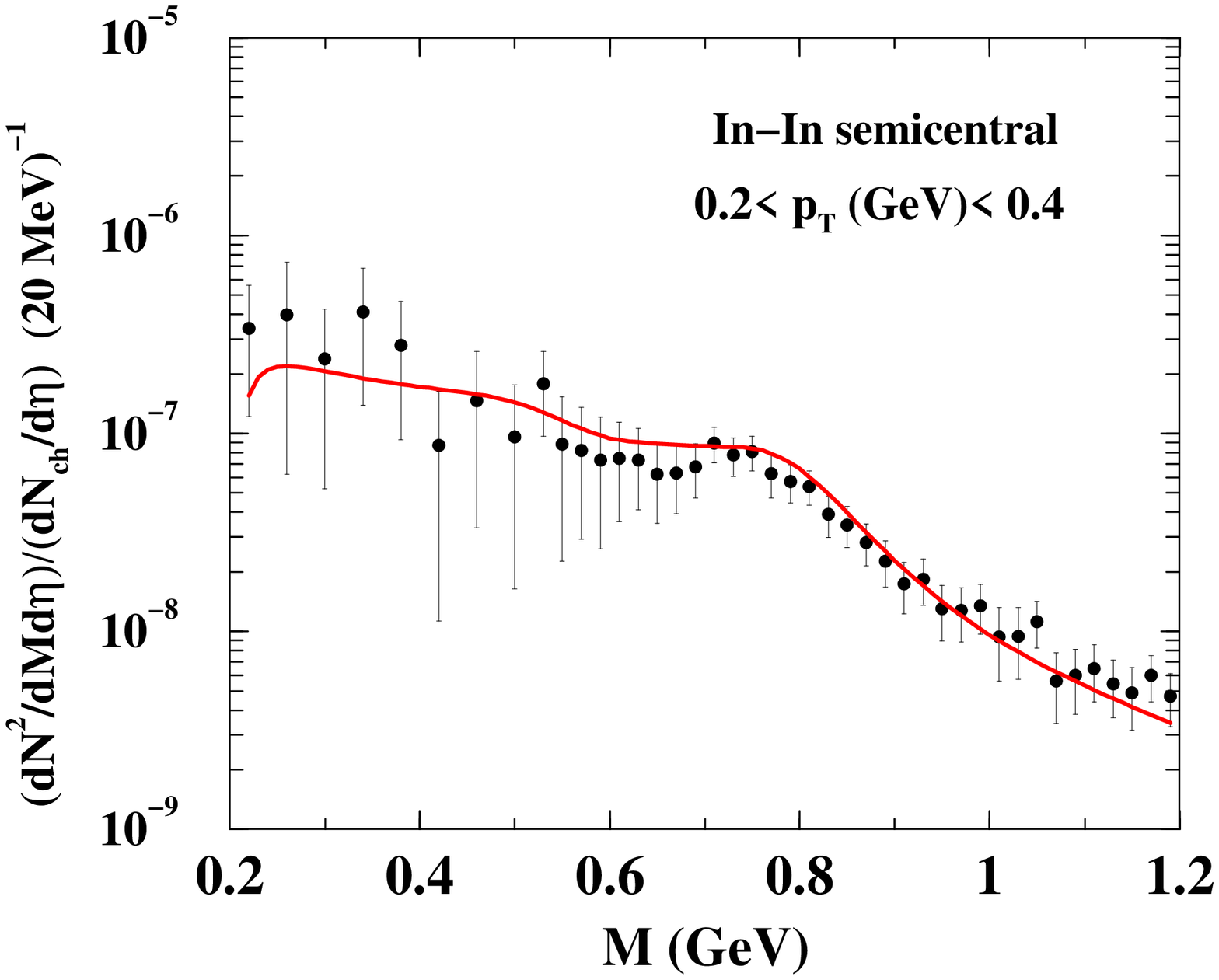}
&\includegraphics[height=5.6cm,width=7.5cm]{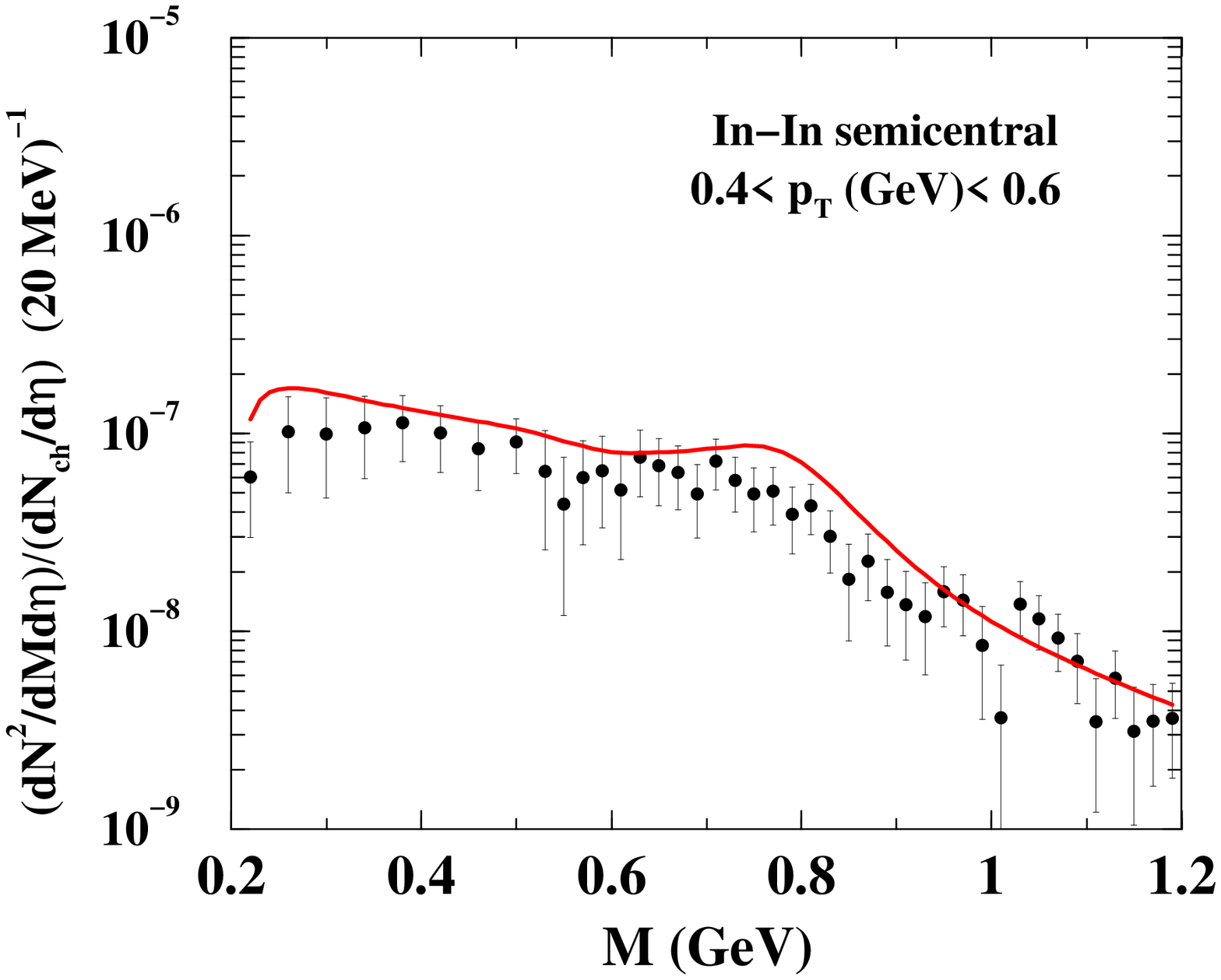}\\
\includegraphics[height=5.6cm,width=7.5cm]{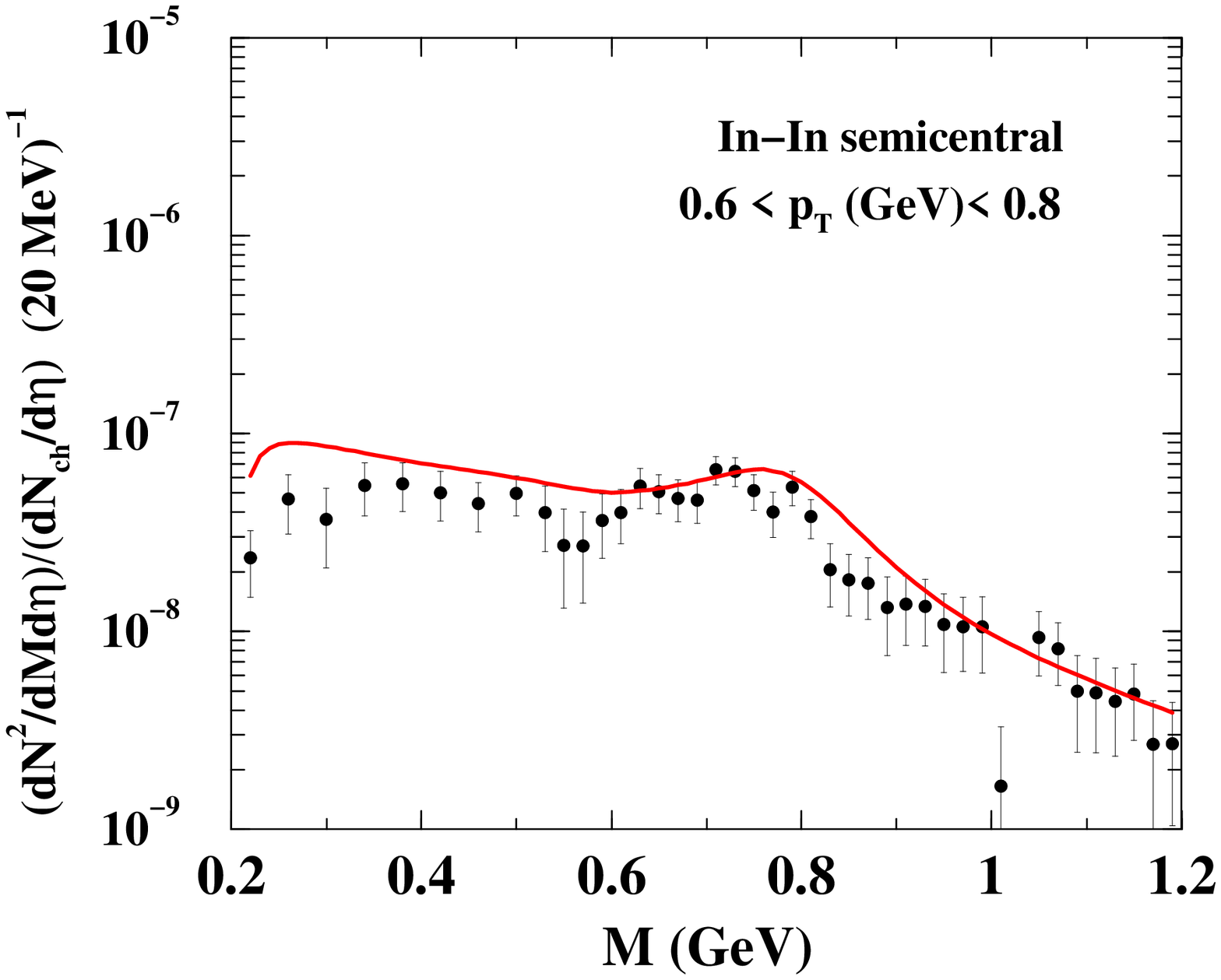}
&\includegraphics[height=5.6cm,width=7.5cm]{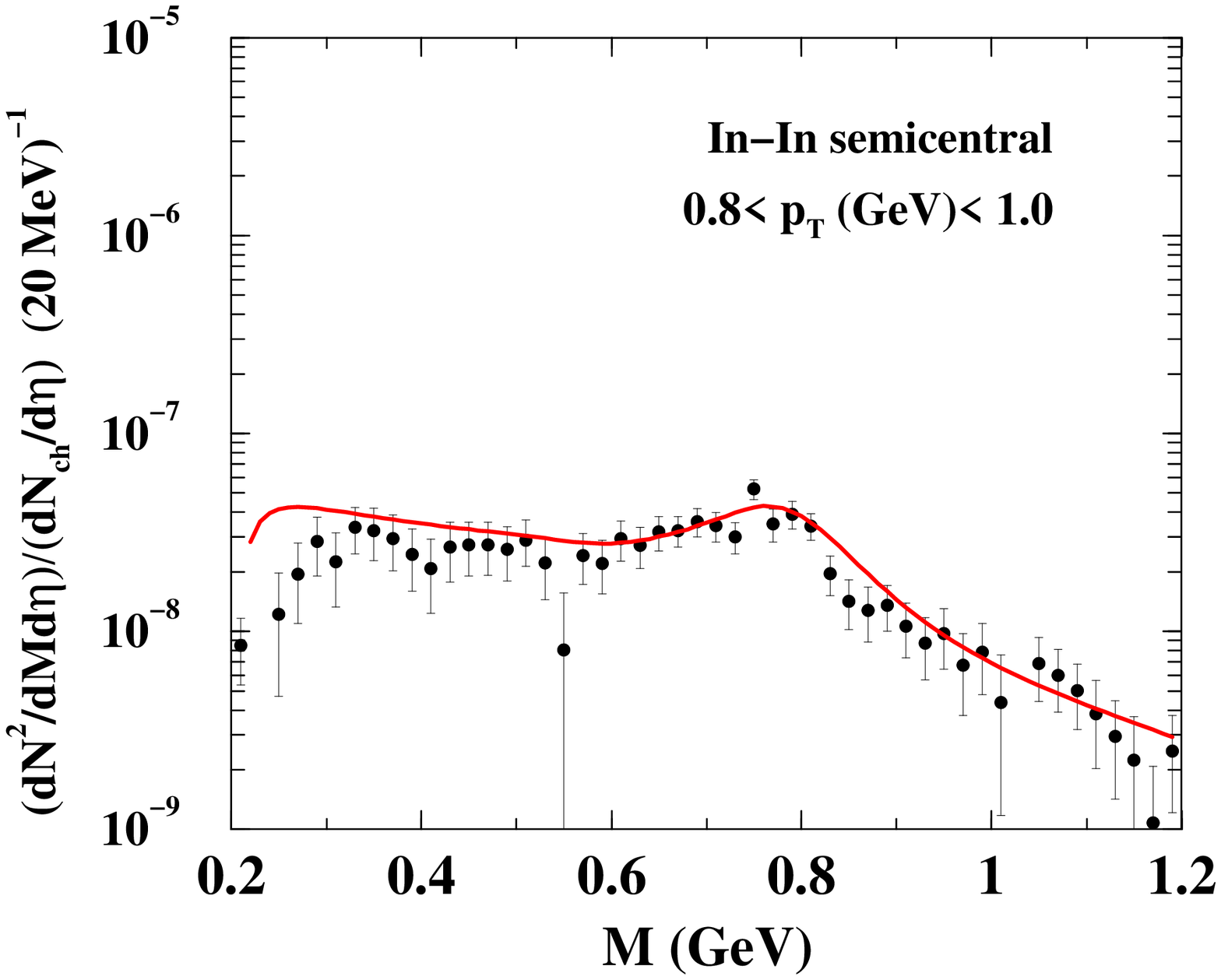}\\
\includegraphics[height=5.6cm,width=7.5cm]{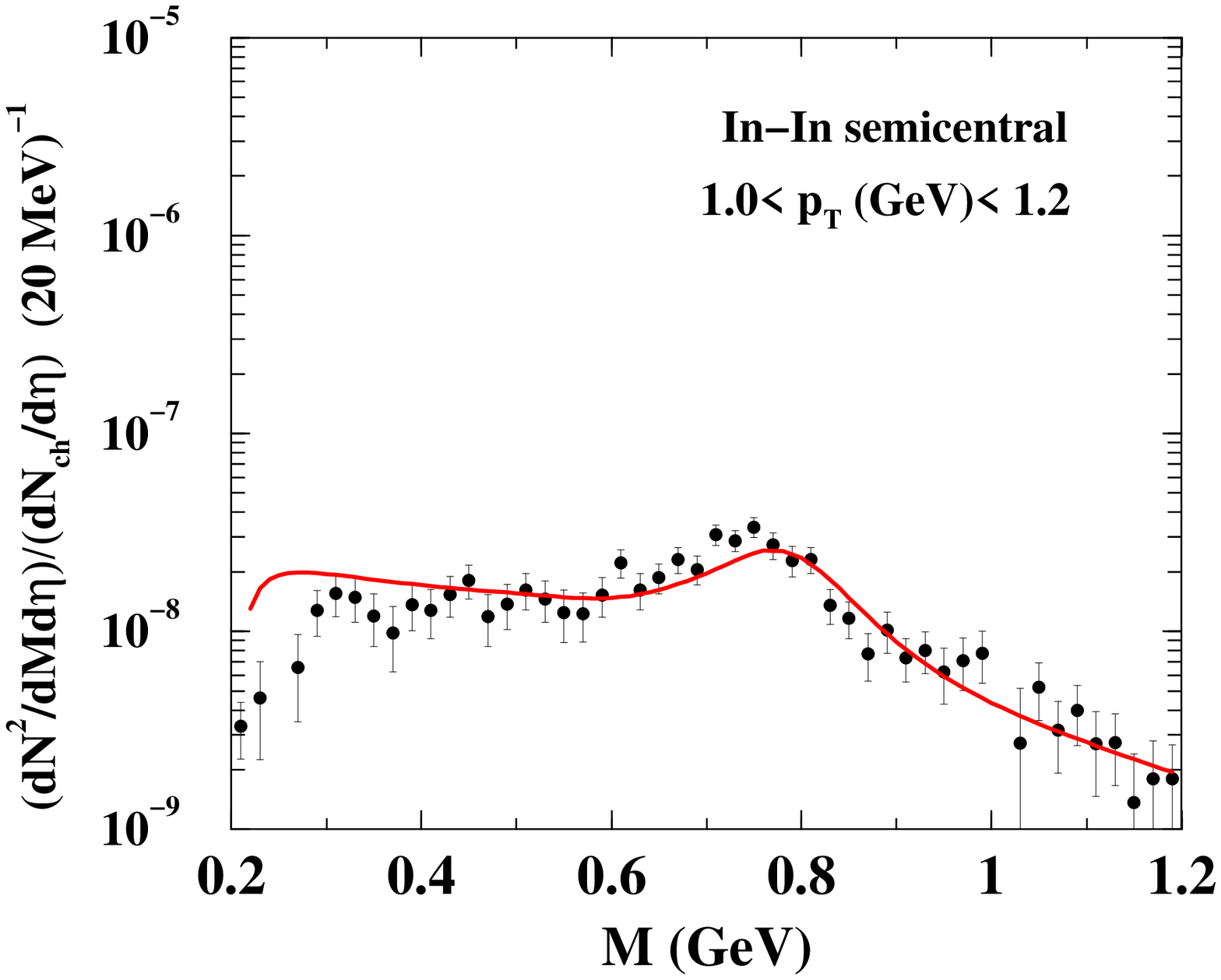}
&\includegraphics[height=5.6cm,width=7.5cm]{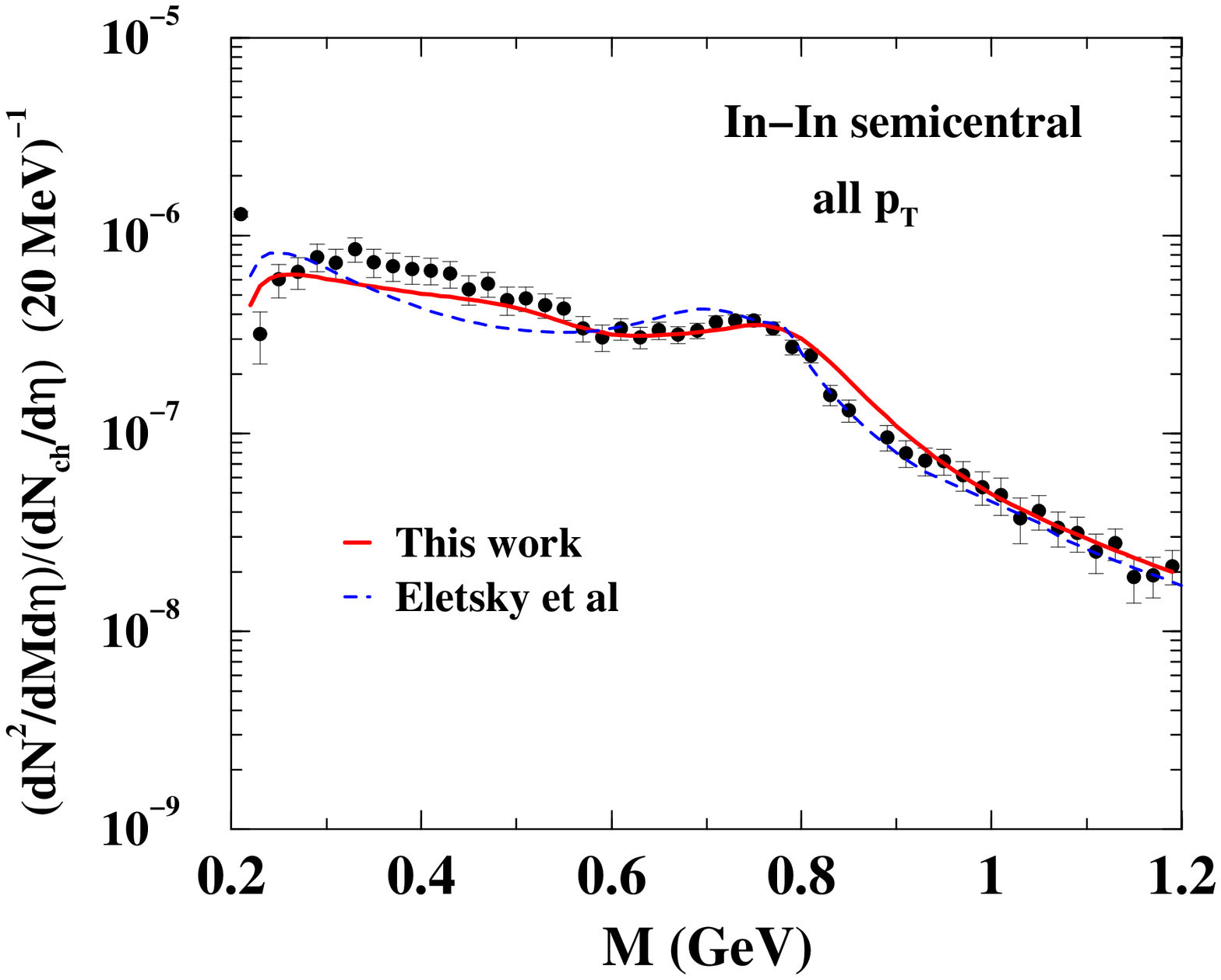}
\label{sps}
\end{tabular}
\end{center}
\caption{Dilepton invariant mass spectra for different $p_T$-bins compared with the NA60
data.}
\label{dNdM}
\end{figure}

Finally, we have obtained the dimuon yield ($dN/dM$) in In-In collisions at
SPS at a center of mass energy of 17.3 AGeV. The initial energy density is taken as 4.5
GeV/fm$^3$ corresponding to a thermalisation time $\tau_i=0.7$ fm. We take the
QGP to hadronic matter transition temperature $T_c=$ 175 MeV and the freeze-out temperature
$T_f=$120 MeV which can reproduce the slope of the hadronic spectra measured by the NA60
Collaboration. In Fig.~(\ref{dNdM}) we have shown the invariant mass spectra for different 
transverse momentum ($p_T$) windows. The theoretical curves agree quite well with the
experimental data~\cite{sanja} for all the $p_T$ ranges. The strong enhancement in the low
$M$ domain is clearly due to the large broadening of the $\rho$ (and $\om$) in the thermal 
medium which comes entirely from the Landau cut in the self-energy diagrams. 
In the last panel we also plot for comparison the spectra calculated in~\cite{jajati}
where the self-energy due to baryons has been evaluated following the approach 
of~\cite{eletsky}.
It is seen that this approach depicted by the dashed curve does not produce
the required enhancement to explain the data in the range $0.35\leq M\leq 0.65$ GeV.

\section{Summary}

The self-energy of $\rho$ and $\omega$ mesons have been computed in nuclear matter at finite
temperature and baryon density. Loop graphs involving mesons, nucleons and 4-star
$N^*$ and $\De$ resonances up to spin 3/2 were calculated using gauge invariant
interactions in the framework of real time thermal field theory to obtain the
correct relativistic expressions for the self-energy. The singularities in
the complex energy plane were analysed and the imaginary part obtained from the
Landau cut contribution. Results for the real and imaginary parts at non-zero
three-momenta for various
values of temperature and baryonic chemical potential were shown for the
individual loop graphs. The spectral function of the 
$\rho$ was observed to undergo a significant modification at and below 
the nominal rho mass which was seen to bring about a large enhancement of
lepton pair yield in this region. After a space-time evolution using relativistic
hydrodynamics, the invariant mass spectra for various $p_T$
windows was found to be in very good agreement with the 
experimental data obtained in In-In collisions
at 17.3 AGeV.

\end{document}